\documentclass[12pt]{article}
\usepackage{amsmath,amssymb}
\usepackage{curves,epic,eepic,graphics}

\begin{document}

\title{\textbf{\large \ Different Prospects of the Universe Depending on Dark Energy and/or a New Kind of Extra Dimension}}
\author{Sheng-Fei Feng$^{1,2}$ \hspace{0.1cm} Yong-Chang Huang$^{1}$ \hspace{0.1cm} Xin Liu$^{1}$ \hspace{0.2cm}Ying-Jie Zhao$^{1}\thanks{%
Corresponding author : yj\_zhao@bjut.edu.cn}$ \\
\\ {\small 1. Institute of Theoretical Physics,}\\ {\small Beijing University of Technology,  Beijing, 100124, China}\\
{\small 2. Department of Physics, Capital Normal University,}\\ {\small Beijing, 100048, China}\\
}
\date{}
\maketitle

\begin{center}
\textbf{Abstract}
\end{center}

Generally making the cosmological scale factor $R$ be a function of
the coordinate of the extra dimension $\sigma $ that is also a function of time $t$, we achieve a new kind
of cosmic acceleration mechanism depending on extra dimension. We give
the constraints on $\sigma $ under four different prospects of the universe, and indicate that dark energy is not required for both the small extra dimension and the accelerating expansion of our universe. This results in this paper show that the accelerating
expansion of our universe may come from both dark energy and the achieved new mechanism here, or the latter alone.
\\
\\
\\
 {\bf PACS Number(s)}: 98.80.-k, 98.80.Jk, 98.80.Qc         

\section{Introduction}

Although the standard cosmology based on General Relativity and inflation
has been remarkably successful, there remain deep puzzles left for theorists
to resolve--- what is the cause of the late-time acceleration of the
universe? For several years, people has put forward a lot of theories to answer the above question. Dark energy, as a famous mechanism, could be interpreted by cosmological constant introduced by Einstein, whose huge difference between the observed value and theoretical  value suggests that it may be a sort  of  strange field. In order to describe dark energy, various theories such as phantom [1], quintessence theories [2], and other forms of exotic matter [3] like the Chaplygin gas [4] (see also Ref.[5] and a more complete list of references) have been proposed. However, the property of dark energy assumed to drive the expansion of the universe can not be probed in any direct experiment.
 
On one hand, at present our knowledge of dark energy still remains insufficient, that is to say,
the fact that we cannot understand the physical essence of dark energy prevents us from researching more accurately;
on the other hand, the acceleration of our universe may be implicated by other new mechanisms  simultaneously,
hence that one can follow a different route and try instead to pin
the acceleration of the universe on the extra dimensions is imperative. Some
theorists modify the Friedmann equations directly [6,7,8]. In this
paper, we are aimed to study this topic using the Kaluza-Klein
formalism where the fifth coordinate is non-compact. In the
following, we will argue how to extend the FRW metrics in section 2. And we derive the Einstein's equations in section 3 and discuss the four different prospects of the universe in section 4, 5, 6, 7. Moreover, we also show that in the limit of zero proper matter-energy density a small extra dimension emerges naturally in section 8. Finally, we make a conclusion in section 9.

\section{Derivation of the Metrics}

The FRW metrics were derived from the following geometrical
structure of space
\begin{equation}
x_{1}^{2}+x_{2}^{2}+x_{3}^{2}+x_{4}^{2}=R^{2},
\end{equation}%
where $x_{1,}x_{2},x_{3},x_{4}$ are all spacial coordinates. In
the calculation the scale factor $R$ was treated as a constant, as
one obtains the metrics, one let $R$ be a function of time $t$. It
seems inconsistent on logic, if $R$ is a function of time it
should be the function from the right beginning. So we think the
fault of the FRW metrics is just coming from this. If there need
modifying, one can start from this. Of course, one can attribute
the fault of the FRW model to the hypothesis that the universe is
homogeneous and isotropic, and think the spacial part of our
universe is
described by $\frac{x_{1}^{2}+x_{2}^{2}+x_{3}^{2}}{R}+\frac{x_{4}^{2}}{L}=1$%
,where $R$ and $L$ are constants [9]. Now, we retain the hypothesis, but let
$R$ be a function of another parameter. However, we find that it is naive if
one only treats $R$ as a function of time from the very beginning
calculation, then one claims that he gets new metrics. In fact, we have
tried to do this, finally, we found the metrics that we had obtained is
faulted as the FRW metrics. On the other hand, more and more theorists
believe that the extra dimensions is existent. So if one wants to describe
the universe, whose metrics must have the terms which come from the extra
dimensions. Based on these reasons, we think that $R$ should be a function
of extra dimension, here, for convenience, we consider the 5D spacetime,
i.e. $Eq.(1)$ should be rewritten as the following form
\begin{equation}
x_{1}^{2}+x_{2}^{2}+x_{3}^{2}+x_{4}^{2}=R^{2}(\sigma ),
\end{equation}%
where $\sigma $ is the extra spacial coordinates. This equation represents a
4D hypersurface which embeds in the 5D Euclidean space. We use intrinsic
coordinates to this hypersurface

\begin{subequations}
\begin{align}
x_{1}& =R(\sigma)\sin \alpha \sin \theta \sin \varphi , \\
x_{2}& =R(\sigma)\sin \alpha \sin \theta \cos \varphi , \\
x_{3}& =R(\sigma)\sin \alpha \cos \theta , \\
x_{4}& =R(\sigma)\cos \alpha .
\end{align}
The metrics of this 4D hypersurface can be given by the the metrics of 5D
Euclidean space.

\end{subequations}
\begin{equation}
(dS_{E})^{2}=(dx_{1})^{2}+(dx_{2})^{2}+(dx_{3})^{2}+(dx_{4})^{2}+(d%
\sigma)^{2}
\end{equation}
The idea that matter in four dimensions (4D) can be explained from a 5D
Ricci-flat $(R_{AB}=0)$ Riemannian manifold is a consequence of the
Campbell's theorem. It says that any analytic $N$-dimensional Riemannian
manifold can be locally embedded in a $(N+1)$-dimensional Ricci-flat
manifold. Substituting $Eqs.(3a-3d)$ into $Eq.(4)$, we can obtain the
following metrics

\begin{equation}
(dS_{E})^{2}=R^{2}(\sigma)\{\frac{dr^{2}}{1-r^{2}}+r^{2}(d\theta ^{2}+\sin
^{2}\theta d\varphi ^{2})\}+[1+R^{\prime 2}(\sigma)]d\sigma^{2}
\end{equation}
where we have denoted the function $\sin \alpha $ as $r$. After considering
the time, the spacetime metrics can be obtained as
\begin{equation}
dS^{2}=dt^{2}-R^{2}(\sigma)\{\frac{dr^{2}}{1-r^{2}}+r^{2}(d\theta ^{2}+\sin
^{2}\theta d\varphi ^{2})\}-[1+R^{\prime 2}(\sigma)]d\sigma^{2}.
\end{equation}

Our space may have negative curvature, in this case, the $Eq.(2)$ can be
rewritten as
\begin{equation}
x_{1}^{2}+x_{2}^{2}+x_{3}^{2}-x_{4}^{2}=-R^{2}(\sigma).
\end{equation}
\ This equation represents a 4D hypersurface which has negative curvature
and immersed in a $5$D pseudo Euclidean flat space with metrics
\begin{equation}
(dS_{p})^{2}=(dx_{1})^{2}+(dx_{2})^{2}+(dx_{3})^{2}-(dx_{4})^{2}+(d%
\sigma)^{2}.
\end{equation}
We use intrinsic coordinates to this hypersurface

\begin{subequations}
\begin{align}
x_{1}& =R(\sigma)\sinh \alpha \sin \theta \sin \varphi , \\
x_{2}& =R(\sigma)\sinh \alpha \sin \theta \cos \varphi , \\
x_{3}& =R(\sigma)\sinh \alpha \cos \theta , \\
x_{4}& =R(\sigma)\cosh \alpha .
\end{align}
The spatial line-element on this hypersurface is obtained from the metrics $%
Eq.(8)$ of the $5$-dimensional pseudo Euclidean flat space in which it is
immersed as

\end{subequations}
\begin{equation}
(dS_{p})^{2}=R^{2}(\sigma)\{\frac{dr^{2}}{1+r^{2}}+r^{2}(d\theta ^{2}+\sin
^{2}\theta d\varphi ^{2})\}+[1-R^{\prime 2}(\sigma)]d\sigma^{2},
\end{equation}
where we have denoted the function $\sinh \alpha $ as $r$. After considering
the time, the spacetime metrics can be obtained as

\begin{equation}
dS^{2}=dt^{2}-R^{2}(\sigma)\{\frac{dr^{2}}{1+r^{2}}+r^{2}(d\theta ^{2}+\sin
^{2}\theta d\varphi ^{2})\}-[1-R^{^{\prime }2}(\sigma)]d\sigma^{2}.
\end{equation}
The two line-elements $(6)$ and $(11)$ can, however, be combined into a
single line-element with the help of a parameter $k$ that takes values $%
k=\pm 1$

\begin{equation}
dS^{2}=dt^{2}-R^{2}(\sigma)\{\frac{dr^{2}}{1-kr^{2}}+r^{2}(d\theta ^{2}+\sin
^{2}\theta d\varphi ^{2})\}-[1+kR^{\prime 2}(\sigma)]d\sigma^{2}.
\end{equation}
It may be noted that if the parameter $k$ is set equal to zero we get the
line-element $Eq.(12)$ as

\begin{equation}
dS^{2}=dt^{2}-R^{2}(\sigma)\{dr^{2}+r^{2}(d\theta ^{2}+\sin ^{2}\theta
d\varphi ^{2})\}-d\sigma^{2},
\end{equation}
which are the metrics of $5$D flat Minkowski spacetime.

Now that we introduce extra dimension into our model, we must
explain why our realistic world is 4-dimensional. To do this, we
take the ansatz that the Standard Model fields are confined to a
3-brane in the extra dimension [10-21]. Considering the brane can
move along the extra dimension, the location of the brane in the
extra dimension should be a function of time. Actually, when $t$
takes a certain value, the large scale state of our universe
should be certain, i.e. if there is a certain value of $t$, the
extra dimension coordinate should be given a certain value too.
Thus, there
must be a relationship between the extra dimension coordinate $\sigma$ and $%
t $. We can express this relation by denoting $\sigma$ be functions of $t$.
We rewrite them as $\sigma(t)$. Since $R(\sigma)$ is the function of $\sigma$%
, now $\sigma$ is a function of $t$, then $R(\sigma)$ finally must be a
function of $t$. We rewrite $R(\sigma)$ as $R(t)$, then the metrics in $%
Eq.(12)$ reduce to $4D$-form and can be rewritten as

\begin{equation}
dS^{2}=[1-\dot{\sigma}^{2}(t)-k\dot{R}^{2}(t)]dt^{2}-R^{2}(t)\{\frac{dr^{2}}{%
1-kr^{2}}+r^{2}(d\theta ^{2}+\sin ^{2}\theta d\varphi ^{2})\}.
\end{equation}

Comparing with Friedmann-Robertson-Walker metrics, we can see, that after
making $R$ be a function of the coordinate of extra dimension, and
considering the coordinate of extra dimension be a function of time, there
appear two new terms in our metrics. These two new terms have new meanings.
They give us freedom to understand the puzzle of the accelerated expansion
of the universe.

 One may argue that the Eq.(14) is an equivalent of FRW metrics
 based on the following reason: if one denotes a new variable
 $\tau$ and lets
 $d\tau^{2}=[1-\dot{\sigma}^{2}(t)-k\dot{R}^{2}(t)]dt^{2}$, then
 the metrics Eq.(14) are exactly FRW metrics. But we point out
 Eq.(14) is not an equivalent of FRW metrics, because
 $g_{00}=1-\dot{\sigma}^{2}(t)-k\dot{R}^{2}(t)$ may be minus, but
 in FRW metrics it is impossible.

\section{Einstein's Equations}

The metrics derived by the aforementioned argument sill
describe a 4D spacetime. We still take the energy-momentum tensor of the 4D
spacetime relativistic perfect fluid form, given by%
\begin{equation}
T_{\mu \nu }=-pg_{\mu \nu }+(p+\rho )U_{\mu }U_{\nu },
\end{equation}%
where $p$ is the isotropic fluid pressure, $\rho $ is the proper
matter-energy density and$\ U_{\nu }(\nu =0,1,2,3)$ is the fluid flow
vector, satisfying $g_{\mu \nu }U^{\mu }U^{\nu }=1$. However, now $g_{\mu
\nu }$ takes the form as following

\begin{equation}
g_{\mu \nu }=\left(
\begin{array}{cccc}
1-\dot{\sigma}^{2}(t)-k\dot{R}^{2}(t) &  &  &  \\
& -\frac{R^{2}(t)}{1-kr^{2}} &  &  \\
&  & -R^{2}(t)r^{2} &  \\
&  &  & -R^{2}(t)r^{2}\sin \theta%
\end{array}
\right) ,
\end{equation}
where $\mu ,\nu =0,1,2,3$.

In the adopted comoving coordinate system we have the components of the
fluid 4-velocity vector $U_{\mu }$ as $U_{1}=U_{2}=U_{3}=0$, Then the
condition $g_{\mu \nu }U^{\mu }U^{\nu }=1$ becomes $U^{0}U^{0}=\frac{1}{1-
\dot{\sigma}^{2}(t)-k\dot{R}^{2}(t)}$, After considering these, the equation
of conservation of the energy-momentum tensor $T_{;\nu }^{\mu \nu }=0$ is

\begin{equation}
-g^{00}p_{,0}+g^{-\frac{1}{2}}[g^{\frac{1}{2}}(p+\rho
)U^{0}U^{0}]_{,0}+\Gamma _{00}^{0}(p+\rho )U^{0}U^{0}=0,
\end{equation}
where

\begin{eqnarray}
\Gamma _{00}^{0} &=&-\frac{\dot{\sigma}\ddot{\sigma}+k\dot{R}\ddot{R}}{1-%
\dot{\sigma}^{2}-k\dot{R}^{2}}, \\
g &=&\frac{\left\vert 1-\dot{\sigma}^{2}-k\dot{R}^{2}\right\vert
R^{6}r^{4}\sin^2 \theta }{1-kr^{2}},
\end{eqnarray}
in which we take the absolute value of $1-\dot{\sigma}^{2}-k\dot{R}^{2}$ to
assure $g$ be a positive number. Calculation show that the usual energy
conservation $\dot{\rho}+3H(\rho +p)=0$ is retained in our model. So, if we
have the equation of state of $p=\omega \rho $, then we can obtain
\begin{equation}
\rho R^{3(\omega +1)}=\frac{3C}{8 \pi G}=const.  \tag{$19'$}
\end{equation}
On the other hand, because many physical processes should satisfy
quantitative causal relation with no-loss-no-gain character
[22,23], e.g., Ref.[24] uses the no-loss-no-gain homeomorphic map
transformation satisfying the quantitative causal relation to gain
exact strain tensor formulas in Weitzenbock manifold. And Eq.
(17) satisfies quantitative causal relation, i.e., changes ( cause
) of some quantities in Eq. (17) must cause relative changes (
result ) of the other quantities in Eq. (17) so that Eq. (17)'s
right side keeps no-loss-no-gain, i.e., maintains zero. Therefore,
Eq.(17) also satisfies a general consistent condition.

From the metrics we can have the Einstein's equations,
first we calculate the Riemannian tensors

\begin{subequations}
\begin{align}
R_{00}& =-3\frac{\ddot{R}}{R}\frac{1-\dot{\sigma}^{2}}{1-\dot{\sigma}^{2}-k%
\dot{R}^{2}}-3\frac{\dot{R}}{R}\frac{\dot{\sigma}\ddot{\sigma}}{1-\dot{\sigma%
}^{2}-k\dot{R}^{2}}, \\
R_{ii}& =\frac{g_{ii}}{R^{2}}\left[ \frac{(R\ddot{R}+2\dot{R}^{2})(1-\dot{%
\sigma}^{2})+\dot{\sigma}\ddot{\sigma}R\dot{R}-2k\dot{R}^{4}}{\left( 1-\dot{%
\sigma}^{2}-k\dot{R}^{2}\right) ^{2}}+2k\right] ,
\end{align}
where $\dot{R}=\frac{\partial R}{\partial t}$, $\ddot{R}=\frac{\partial ^{2}R%
}{\partial ^{2}t}$ and $i=1,2,3$, double $i$ dose not mean summation.
Substituting these equations into the Einstein's equations

\end{subequations}
\begin{equation}
R_{\mu \nu }=8\pi G(T_{\mu \nu }-\frac{1}{2}g_{\mu \nu }T)=8\pi GS_{\mu \nu
},
\end{equation}
where $T=g^{\mu \nu }T_{\mu \nu },$ $S_{\mu \nu }=T_{\mu \nu }-\frac{1}{2}%
g_{\mu \nu }T.$ One can calculate that

\begin{eqnarray}
S_{00} &=&\frac{1}{2}(\rho +3p)g_{00}, \\
S_{ii} &=&\frac{1}{2}(p-\rho )g_{ii},
\end{eqnarray}%
We must notice, where $g_{00}=1-\dot{\sigma}^{2}(t)-k\dot{R}^{2}(t)$, which
is different from FRW model, and $i=1,2,3$, double $i$ dose not mean
summation. Now we have the Einstein's equations in our metrics
\begin{subequations}
\begin{gather}
-3\frac{\ddot{R}}{R}(1-\dot{\sigma}^{2})-3\frac{\dot{R}}{R}\dot{\sigma}\ddot{%
\sigma}=4\pi G(\rho +3p)\left( 1-\dot{\sigma}^{2}-k\dot{R}^{2}\right) ^{2},
\\
\frac{(R\ddot{R}+2\dot{R}^{2})(1-\dot{\sigma}^{2})+\dot{\sigma}\ddot{\sigma}R%
\dot{R}-2k\dot{R}^{4}}{\left( 1-\dot{\sigma}^{2}-k\dot{R}^{2}\right) ^{2}}%
+2k=4\pi G(\rho -p)R^{2}.
\end{gather}%
Substituting $Eq.(24a)$ into $Eq.(24b)$, we have the Friedmann-Equation in
our model

\end{subequations}
\begin{equation}
\dot{R}^{2}=\left( \frac{8\pi G\rho R^{2}}{3}-k\right) \left( 1-\dot{\sigma}%
^{2}-k\dot{R}^{2}\right) .
\end{equation}%
From this equation, and after considering the equation of energy
conservation, we can obtain the relation of $\sigma $ and $R$ as following
\begin{equation}
\dot{\sigma}^{2}=1-k\dot{R}^{2}-\frac{\dot{R}^{2}R^{3\omega +1}}{\left(
C-kR^{3\omega +1}\right) },
\end{equation}
where $\frac{8\pi G\rho R^{3(\omega +1)}}{3}=C$ is a positive
number. The parameter $\sigma $ comes of the extra dimension.
$Eq.(26)$ describes the relation between $\sigma (t)$ and $R(t)$.
If we can write $R(t)$ with an obvious form, then we can obtain
the knowledges about the extra dimension $\sigma (t)$. We hope
that this new parameter can give us freedom to explain the
accelerated expansion of the universe. Take the derivative of
$Eq.(26)$ with respect to time, we have

\begin{equation}
k\dot{R}\ddot{R}+\frac{\dot{R}\ddot{R}R^{3\omega +1}+\frac{3\omega +1}{2}%
\dot{R}^{3}R^{3\omega }}{\left( C-kR^{3\omega +1}\right) }+\frac{k\frac{%
3\omega +1}{2}\dot{R}^{3}R^{6\omega +1}}{\left( C-kR^{3\omega +1}\right)^2 }=-%
\dot{\sigma}\ddot{\sigma}.
\end{equation}
There are three cases about this equation, i.e.

\begin{equation}
\left\{
\begin{array}{l}
\ddot{R}=\frac{3\omega +1}{2}\frac{\dot{R}^{2}R^{3\omega }}{\left(
C+R^{3\omega +1}\right) }+\frac{\left( C+R^{3\omega +1}\right) \dot{\sigma}%
\ddot{\sigma}}{\dot{R}C},\text{ \ \ \ \ \ \ \ \ \ \ \ \ \ }k=-1, \\
\ddot{R}=-\frac{3\omega +1}{2}\frac{\dot{R}^{2}}{R}-\frac{C\dot{\sigma}\ddot{%
\sigma}}{\dot{R}R^{3\omega +1}},\text{ \ \ \ \ \ \ \ \ \ \ \ \ \ \ \ \ \ \ \
\ \ \ \ \ \ \ }k=0, \\
\ddot{R}=-\frac{3\omega +1}{2}\frac{\dot{R}^{2}R^{3\omega }}{\left(
C-R^{3\omega +1}\right) }-\frac{\left( C-R^{3\omega +1}\right) \dot{\sigma}%
\ddot{\sigma}}{\dot{R}C},\text{ \ \ \ \ \ \ \ \ \ \ \ }k=1.%
\end{array}%
\right.
\end{equation}%

In the following space we investigate the constraints on extra dimension $\sigma(t)$ corresponding respectively to four different prospects of the universe.

\section{Big Bounce ($\dot{R}=0, \ddot{R}>0$)}

From $Eq.(25)$ we can obtain
\begin{equation}
R = \sqrt {\frac{{3k}}{{8\pi G\rho }}}, \label{br1}
\end{equation}
or
\begin{equation}
\dot{\sigma}=\pm 1, \quad \ddot{R}=0, \quad   (\textrm{here we drop it because  }\ddot{R}>0)
\end{equation}

In the $Eq.(29)$, $k$ must satisfy $k=1$.
Moreover, according to $Eq.(24a)$ and $p = \omega \rho$, we have the relation
\begin{equation}
\left( {1 + 3 \omega} \right)\left( {1 - {{\dot \sigma }^2}} \right)\rho < 0.
\end{equation}
When $\omega > -1/3 $ we get
\begin{equation}
{\dot \sigma }>1 \qquad \textrm{or} \qquad {\dot \sigma }<-1,
\end{equation}
or  $\omega < -1/3 $ we get
\begin{equation}
-1 < {\dot \sigma } <1.
\end{equation}
And the extra dimension $\sigma(t)$ is represented from $Eq.(24b)$ as
\begin{equation}
 \dot \sigma  =  \pm \sqrt {1 - \frac{{2R^{3\omega + 2}\ddot R}}{{3C\left( {1  - \omega}\right) - 4 R^{3\omega + 1 }}}}.
\end{equation}
Thus only the curvature of the space is positive the Big Bounce will happen, and the critical density and cosmological scale factor  corresponding to the time $t_0$ can be obtained from $Eqs$.(\ref{br1}) and ($19'$),
\begin{equation}
R(t_0) = C^{\frac{1}{3\omega+1}}, \quad \rho(t_0) =  \frac{3}{8 \pi G C^{\frac{2}{3\omega+1}}}.
\end{equation}

\section{Big Rip ($\dot{R}>0, \ddot{R}\geq0$)}

Because of the generality of $\sigma $, there is always $\ddot{R}\geq0$, as
long as one takes an appropriate form of the function $\sigma (t)$. Now, we
consider the constraints on $\sigma (t)$, which make $\ddot{R}\geq0$. Therefore,
using $\ddot{R}\geq0$ in $Eq.(28)$, we have the inequations about $\dot{\sigma}$ $(\omega >-1/3)$

\begin{equation}
\left\{
\begin{array}{l}
\frac{d{\dot\sigma}^2}{dt} \geqslant -\frac{{{{\left(3\omega  + 1\right)}}C{{\dot R}^3}{R^{3\omega }}}}{{{{\left( {C +{R^{3\omega  + 1}}} \right)}^2}}},
\text{\ \ \ \ \ \ \ \ \ \ \ \ \ \ }k=-1, \\
\frac{d{\dot\sigma}^2}{dt}\leqslant - \frac{{\left(3\omega  + 1\right)}{\dot R}^3 R^{3\omega}}{C},
\text{ \ \ \ \ \ \ \ \ \ \ \ \ \ \ }k=0, \\
\frac{d{\dot\sigma}^2}{dt}\geqslant - \frac{{{\left(3\omega  + 1\right)}C{{\dot R}^3}{R^{3\omega }}}}{{{{\left( {{R^{3\omega  + 1}} -C} \right)}^2}}},\text{ \ \ \ \ \ \ \ \ \ \ \ \ }k=1,  R>C^{\frac{1}{3\omega +1}}.
\end{array}%
\right.
\end{equation}%
According to above conditions, it is clear that the growing $R(t)$ must be larger than any constant (see $Eq.(28)$), hence only  $R>C^{\frac{1}{3\omega +1}}$ can the Big Rip happen in the universe having positive curvature  $(k=1)$.
But when $\omega<-1/3$ the inequations are
\begin{equation}
\left\{
\begin{array}{l}
\frac{d{\dot\sigma}^2}{dt} \geqslant - \frac{{{\left(3\omega  + 1\right)}C{{\dot R}^3}{R^{3\omega }}}}{{{{\left( {C +{R^{3\omega  + 1}}} \right)}^2}}},
\text{\ \ \ \ \ \ \ \ \ \ \ \ \ \ }k=-1, \\
\frac{d{\dot\sigma}^2}{dt} \leqslant - \frac{{\left(3\omega  + 1\right)}{\dot R}^3 R^{3\omega}}{C},
\text{ \ \ \ \ \ \ \ \ \ \ \ \ \ \ \ }k=0, \\
\frac{d{\dot\sigma}^2}{dt} \leqslant - \frac{{{\left(3\omega  + 1\right)}C{{\dot R}^3}{R^{3\omega }}}}{{{{\left( {{R^{3\omega  + 1}-C}} \right)}^2}}},\text{ \ \ \ \ \ \ \ \ \ \ \ \ \ }k=1,  R>C^{\frac{1}{3\omega +1}}.%
\end{array}%
\right.
\end{equation}%
If the above conditions are satisfied, the universe maintains its infinitely accelerating inflation so that the Big Rip will occur.

\section{No Big Rip ($\dot{R}>0, \ddot{R}<0$)}
Similarly, according to $Eq.(28)$, the extra dimension $\sigma(t)$ is restricted  $(\omega>-1/3)$
\begin{equation}
\left\{
\begin{array}{l}
\frac{d{\dot\sigma}^2}{dt} < -  \frac{{{\left(3\omega  + 1\right)}C{{\dot R}^3}{R^{3\omega }}}}{{{{\left( {C +{R^{3\omega  + 1}}} \right)}^2}}},
\text{\ \ \ \ \ \ \ \ \ \ \ \ \ \ }k=-1, \\
\frac{d{\dot\sigma}^2}{dt} > -  \frac{{\left(3\omega  + 1\right)}{\dot R}^3 R^{3\omega}}{C},
\text{ \ \ \ \ \ \ \ \ \ \ \ \ \ \ \ }k=0, \\
\frac{d{\dot\sigma}^2}{dt} < - \frac{{{\left(3\omega  + 1\right)}C{{\dot R}^3}{R^{3\omega }}}}{{{{\left( {{R^{3\omega  + 1}}-C} \right)}^2}}},\text{ \ \ \ \ \ \ \ \ \ \ \ \ \ }k=1, R>C^{\frac{1}{3\omega +1}}.%
\end{array}%
\right.
\end{equation}%
 But when $\omega<-1/3$ the inequations are
\begin{equation}
\left\{
\begin{array}{l}
\frac{d{\dot\sigma}^2}{dt} < -  \frac{{{\left(3\omega  + 1\right)}C{{\dot R}^3}{R^{3\omega }}}}{{{{\left( {C +{R^{3\omega  + 1}}} \right)}^2}}},
\text{\ \ \ \ \ \ \ \ \ \ \ \ \ \ \ }k=-1, \\
\frac{d{\dot\sigma}^2}{dt}  > -  \frac{{\left(3\omega  + 1\right)}{\dot R}^3 R^{3\omega}}{C},
\text{ \ \ \ \ \ \ \ \ \ \ \ \ \ \ \ }k=0, \\
\frac{d{\dot\sigma}^2}{dt} > -  \frac{{{\left(3\omega  + 1\right)}C{{\dot R}^3}{R^{3\omega }}}}{{{{\left( {{R^{3\omega  + 1}}-C} \right)}^2}}},\text{ \ \ \ \ \ \ \ \ \ \ \ \ \ }k=1,  R>C^{\frac{1}{3\omega +1}}.%
\end{array}%
\right.
\end{equation}%
Because this situation is the expansion of decreasing accaleration, there is no Big Rip.
\section{Big Crunch ($\dot{R}<0, \ddot{R}<0$)}

According to $Eq.(28)$ and when $\omega >-1/3$, the extra dimension $\sigma(t)$ is restricted as
\begin{equation}
\left\{
\begin{array}{l}
\frac{d{\dot\sigma}^2}{dt} > -  \frac{{{\left(3\omega  + 1\right)}C{{\dot R}^3}{R^{3\omega }}}}{{{{\left( {C +{R^{3\omega  + 1}}} \right)}^2}}},
\text{\ \ \ \ \ \ \ \ \ \ \ \ \ \ }$$k=-1, \\
\frac{d{\dot\sigma}^2}{dt} < -  \frac{{\left(3\omega  + 1\right)}{\dot R}^3 R^{3\omega}}{C},
\text{ \ \ \ \ \ \ \ \ \ \ \ \ \ \ \ }k=0, \\
\frac{d{\dot\sigma}^2}{dt} > -  \frac{{{\left(3\omega  + 1\right)}C{{\dot R}^3}{R^{3\omega }}}}{{{{\left( {C -{R^{3\omega  + 1}}} \right)}^2}}},\text{ \ \ \ \ \ \ \ \ \ \ \ \ \ }k=1, R < C^{\frac{1}{3\omega +1}}. %
\end{array}%
\right.
\end{equation}
In the same way, it is not difficult to show the declining $R(t)$ must be smaller than any constant, therefore only $R<C^{\frac{1}{3\omega +1}}$  can the Big Crunch process continuously works in the universe with positive curvature $(k=1)$. But when $\omega<-1/3$ the inequations are
\begin{equation}
\left\{
\begin{array}{l}
\frac{d{\dot\sigma}^2}{dt} > -  \frac{{{\left(3\omega  + 1\right)}C{{\dot R}^3}{R^{3\omega }}}}{{{{\left( {C +{R^{3\omega  + 1}}} \right)}^2}}},
\text{\ \ \ \ \ \ \ \ \ \ \ \ \ \ }k=-1, \\
\frac{d{\dot\sigma}^2}{dt} < -  \frac{{\left(3\omega  + 1\right)}{\dot R}^3 R^{3\omega}}{C},
\text{ \ \ \ \ \ \ \ \ \ \ \ \ \ \ \ }k=0, \\
\frac{d{\dot\sigma}^2}{dt} < -  \frac{{{\left(3\omega  + 1\right)}C{{\dot R}^3}{R^{3\omega }}}}{{{{\left( {C -{R^{3\omega  + 1}}} \right)}^2}}},\text{ \ \ \ \ \ \ \ \ \ \ \ \ \ }k=1,  R<C^{\frac{1}{3\omega +1}}.%
\end{array}%
\right.
\end{equation}%
If the above conditions are satisfied, the universe will maintain its accelerating shrink and the Big Crunch will occur until it reaches a singularity.
\section{A Limit Case}
Now we consider a limit case the density of our universe $\rho \rightarrow 0$, that is to say, if $\frac{C}{R^{3\omega +1}}\rightarrow 0$, $Eq.(26)$
will give the extremum of $\dot{\sigma}^{2}$ when $k=\pm 1$, i.e.
\begin{equation}
\underset{\frac{C}{R^{3\omega +1}}\rightarrow 0}{\lim }\dot{\sigma}^{2}=%
\underset{\frac{C}{R^{3\omega +1}}\rightarrow 0}{\lim }\left[ 1-k\dot{R}^{2}-%
\frac{\dot{R}^{2}}{\left( \frac{C}{R^{3\omega +1}}-k\right) }\right] =1,
\end{equation}%
which means that now value (now R is enough big) of the speed of our
universe which can be regarded as a 4D hypersurface embeded in 5D
bulk spacetime will tend to a constant. Further,
$\dot{\sigma}^{2}=1$, i.e. $\dot{\sigma}=\pm 1$, where the
positive sign denotes that the size of the exra dimension is
becoming biger and the minus sign denotes that the size is
becoming smaller, the becoming small size is agreeable with
 the small extra dimension. Consequently, we deduce not only
the small extra dimension but also the accelerating expansion of our universe
 not needing dark energy.

Finally, we can give the redshift in our model
\begin{equation}
z=\alpha\frac{R(t_{0})}{R(t_{e})}-1,
\end{equation}
where
$\alpha=\sqrt{\frac{1-\dot{\sigma}^{2}(t_{e})-k\dot{R}^{2}(t_{e})}{1-\dot{\sigma}^{2}(t_{0})-k\dot{R}^{2}(t_{0})}}$.
$Eq.(32)$ can be rewritten as
\begin{equation}
z=\alpha[1+H_{0}(t_{0}-t_{e})+\left(1+\frac{q_{0}}{2}\right)H^{2}_{0}(t_{0}-t_{e})^{2}+\ldots]-1,
\end{equation}
where $H_{0}=\frac{\dot{R}(t_{0})}{R(t_{0})}$ is Hubble constant
and $q_{0}=-\frac{{\ddot R(t_{0})}}{R(t_{0})H^{2}_{0}}$ is
deacceleration parameter. For diferent $\alpha$, we can obtain different redshift
compared with FRW model, which may make us have more freedom to fit with astronomic
observation results.

\begin{section}
{Summary and Conclusion}
\end{section}

To explain the late-time acceleration of our universe, we need
new mechanism, because the FRW model encounters blank wall in the
new data of observations. In order to get a new mechanism, the
most convenient way is to modify the FRW model. After aboratively
analyzing the FRW model, we show that we can modify the FRW
model by introducing an extra dimension of space $\sigma $, and
making the cosmological scale factor $R$ be a function of the
extra dimension, $R\rightarrow R(\sigma )$. We find the manifold
of space of our universe can be described by the $Eq.(2)$, because the
observations show that our universe is homogeneous and isotropic.
Basing on this considering, we obtain the metrics $Eq.(12)$. As a
modification of the FRW model, it should be able to go back to FRW
model, so we let the coordinate of the extra dimension $\sigma $
be a function of time. Then the metrics $Eq.(12)$ turn to
$Eq.(14)$. Subsequently, we considered the equation of
conservation of the energy-momentum tensor in our metrics, and we
show that the equation of conservation of the energy-momentum in
FRW model can be remained in our model.

We further derive Einstein's equations. In nature, Einstein's
equations describe the relation $Eq.(26)$ between $\sigma (t)$ and
$R(t)$. If we can write $R(t)$ with an obvious form, then we can
obtain the knowledge about the extra dimension $\sigma (t)$.
Observations show that the expansion of our universe is
accelerated, which imposes some constraints to $\sigma (t)$. Here
we give the constraints from $Eq.(33)$ to $Eq.(40)$ on $\sigma (t)$ corresponding to four different prospects of our universe.  As we see in the limitation $\rho \rightarrow 0$, after introducing the extra dimension $\sigma (t)$, we can obtain
not only an accelerated expansion of the universe with dark
energy but also the now small extra dimension, even with the latter alone. Therefore, the achieved results in this letter represent that the accelerating
expansion of our universe may come from the achieved new mechanism.
This Letter's research can also be applied to relative investigations about Refs.[25,26] etc.\\

ACKNOWLEDGMENTS\\

This work is supported by National Natural Science Foundation of China
(Nos.11275017 and 11173028).
\\
\\


\begin{thebibliography}{9}
\bibitem{} R. R. Calwell, Phys. Lett. B 545, 23 (2002); R. R. Caldwell, M.
Kamionkowski and N. N. Weinberg, Phys. Rev. Lett. 91, 071301 (2003); J. M.
Cline, S. Y. Jeon and G. D. Moore, Phys. Rev. D 70, 043543 (2004).

\bibitem{} R. R. Caldwell, R. Dave and P. J. Steinhardt, Phys. Rev. Lett.
80, 1582 (1998); P. J. E. Peebles and A. Vilenkin, Phys. Rev. D 59, 063505
(1999); P. J. Steinhardt, L. M. Wang and I. Zlatev, Phys. Rev. D 59, 123504
(1999); M. Doran and J. Jackel, Phys. Rev. D 66, 043519 (2002); A. R.
Liddle, P. Parsons and J. D. Barrow, Phys. Rev. D 50, 7222 (1994).

\bibitem{} H. Wei, R. G. Cai and D. F. Zeng, Class. Quant. Grav. 22, 3189
(2005) [arXiv:hep-th/0501160].

\bibitem{} A. Y. Kamenshchik, U. Moschella and V. Pasquier, Phys. Lett. B
511 (2001) 265.

\bibitem{} V. Sahni and A. Starobinsky, arXiv:astro-ph/0610026.

\bibitem{} Katherine Freese. Nuclear Physics B (Proc. Suppl.) 124 (2003)
50-54.

\bibitem{} Katherine Freese, Matthew Lewis. Phys. Lett. B540 (2002) 1-8.

\bibitem{} Katherine Freese. New Astron. Rev. 49 (2005) 103-109.

\bibitem{} Abdussattar (Banaras Hindu University). Int. J. Mod. Phys. D12
(2003) 1603-1614.
\bibitem{} V.A. Rubakov, M.E. Shaposhnikov, Phys. Lett. B
125 (1983) 136, 139.
\bibitem{} K. Akama, Lect. Notes Phys. 176 (1982) 267
[hep-th/0001113].
\bibitem{} M. Visser, Phys. Lett. B 159 (1985) 22
[hep-th/9910093].
\bibitem{} M. Pavsic, Phys. Lett. A 116 (1986) 1
[gr-qc/0101075].
\bibitem{} G.W. Gibbons, D.L. Wiltshire, Nucl. Phys. B 287
(1987) 717 [hep-th/0109093].
\bibitem{} N. Arkani-Hamed, S. Dimopoulos, G.R.
Dvali, Phys. Lett. B 429 (1998) 263.
\bibitem{} A.Lukas, B.A. Ovrut, K.S.
Stelle, D. Waldram,Phys. Rev. D 59 (1999) 086001.
\bibitem{} N. Arkani-Hamed,
S. Dimopoulos, G.R. Dvali,Phys. Rev. D 59 (1999) 086004.
\bibitem{} I.
Antoniadis, N. Arkani-Hamed, S. Dimopoulos, G.R. Dvali, Phys.
Lett. B 436 (1998) 257.
\bibitem{} M. Gogberashvili, Mod. Phys. Lett. A 14
(1999) 2025.
\bibitem{} ANIRUDH PRADHAN and I. AOTEMSHI, Int. J. Mod. Phys. D 11 (2002) 1639.
\bibitem{} L. Randall, R.
Sundrum, Phys. Rev. Lett. 83 (1999) 3370, 4690.

\bibitem{} Yong-Chang Huang et al, Mod. Phys. Lett., A21: 1107 (2006) ; Leng Liao and Yong-Chang Huang, Phys. Rev., D75 (2007) 025025.

\bibitem{} Yong-Chang Huang, Chen-Xin Yu, Phys. Rev., D75 (2007) 044011;
Chen-Xin Yu, Yong-Chang Huang, Phys. lett. B647, 49(2007);
Yong-Chang Huang and Fang-Fang Yuan, Journal of High Energy Physics,
03(2011)029.

\bibitem{} Yong-Chang Huang and B. L. Lin, Phys. Lett. , A299 (2002) 644.
\bibitem{} P. Huang and Yong-Chang Huang, Stability of the holographic description of the Universe, European Physical Journal C, (2013) 73:2366.
\bibitem{} Q. Zhang and Yong-Chang Huang, DBI Potential, DBI Inflation Action and General Lagrangian Relative to Phantom, K-essence and Quintessence, Journal of Cosmology and Astroparticle Physics, 11(2011)050.
\end{thebibliography}
\end{document}